\documentclass[10pt,conference,compsocconf]{IEEEtran}
\IEEEoverridecommandlockouts
\usepackage{times}
\usepackage{caption}
\usepackage{multirow}
\usepackage{url}
\ifCLASSINFOpdf
  \usepackage[pdftex]{graphicx}
  \DeclareGraphicsExtensions{.pdf,.jpeg,.png}
\else
\fi

\usepackage{tikz}
\usepackage{subcaption}
\usepackage{pgfplots}
\pgfplotsset{compat=1.18}

\begin{document}
\pagenumbering{gobble}
\title{\textbf{\Large System for Measurement of Electric Energy\\ Using Beacons with Optical Sensors and LoRaWAN Transmission\\[-1.5ex] }}


\author{\IEEEauthorblockN{~\\[-0.4ex]\large \L{}ukasz Marcul \\ and Mateusz Brzozowski\\[0.3ex]\normalsize}
\IEEEauthorblockA{OneMeter Ltd.\\ul. Dobrza\'{n}skiego 3\\
20-262 Lublin, Poland\\
Email: \tt \{Lukasz.Marcul,\\ \tt Mateusz.Brzozowski\}@onemeter.com}
\and
\IEEEauthorblockN{~\\[-0.4ex]\large Artur Janicki\\[0.3ex]\normalsize}
\IEEEauthorblockA{Warsaw University of Technology\\
ul. Nowowiejska 15/19\\
00-665 Warsaw, Poland\\
%OneMeter Ltd., Lublin, Poland\\
Email: {\tt Artur.Janicki@pw.edu.pl}}
\thanks{This work was supported by the National Centre for~Research and Development within the Smart Growth Operational Programme (agreement No.~POIR.01.02.00-00-0352/16-00).}
}



\maketitle

\begin{abstract}
%\boldmath
In this article, we present the results of experiments with finding an efficient radio transmission method for an electric energy measurement system called OneMeter 2.0. This system offers a way of collecting energy usage data from beacons attached to regular, non-smart meters. 
In our study, we compared several low power wide area network (LPWAN) protocols, out of which we chose the LoRaWAN protocol. We verified the energy consumption of a LoRa-based transmission unit, as well as the transmission range between network nodes in urban conditions. We discovered that LoRaWAN-based transmission was highly energy-efficient and offered decent coverage, even in a difficult, dense urban environment.
\end{abstract}



% \fi
%
% For peer review papers, this IEEEtran command inserts a page break and
\IEEEpeerreviewmaketitle

\section{Introduction}
Real-time energy consumption monitoring is a must in our times, as the economic and societal costs of energy production are growing. The European Commission required their member countries to equip at least 80\% of their electricity customers with intelligent metering systems by 2020~\cite{EuDirective2012}. This was supposed to lead to the creation of smart power grids~\cite{Morello2017Smart}, allowing for easy monitoring and managing of country-based and EU-based power consumption. 

The process of installing smart meters is very costly and time-consuming, so it is no wonder that most of the EU countries, as described in the next section, did not meet the above deadline. Therefore, to improve the deployment process of smart metering, we proposed a system called OneMeter 2.0~\cite{Brzozowski2019Enhancing}, which used energy-efficient beacons, usually with optical sensors, communicating, e.g., via the IEC 65056-21 protocol. The system adds intelligent functionality to existing, popular, non-smart, electronic meters, called Automated Meter Reading (AMR)~\cite{Garcia2017Power}, equipped with an optical port or even only with a blinking LED diode, without a need to install smart meters at all.

In this study, we focus on finding an efficient radio communication protocol that would be used for communication between beacons and the cloud. We will choose a suitable low power wide area network (LPWAN) protocol and then verify its energy efficiency and radio coverage offered.

Our paper is structured as follows: in Section~\ref{sec:SmartMetering}, we will briefly describe the problem of smart metering deployment. Next, in Section~\ref{sec:OMProject} we will describe the OneMeter 2.0 system, including the proposed usage of LPWAN-based communication system. Next, we will describe our experiments (Section~\ref{sec:Experiments}), followed by their results, presented in Section~\ref{sec:Results}. Finally, we will conclude in Section~\ref{sec:Conclusions} with a plan for the future of our work.

\begin{figure}[!t]
\centering
\includegraphics[width=0.7\linewidth]{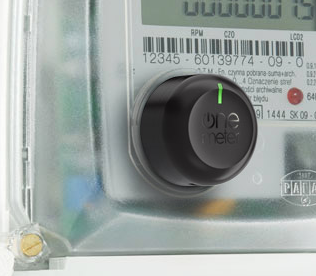}
\caption{OneMeter beacon attached to IEC 62056-21 interface of electricity meter.}
\label{fig:beacon}
\end{figure}

\section{Smart Metering Challenges}
\label{sec:SmartMetering}
\subsection{Smart Metering in Europe}

%OM system
\begin{figure*}[!t]
\centering
\includegraphics[width=0.9\linewidth]{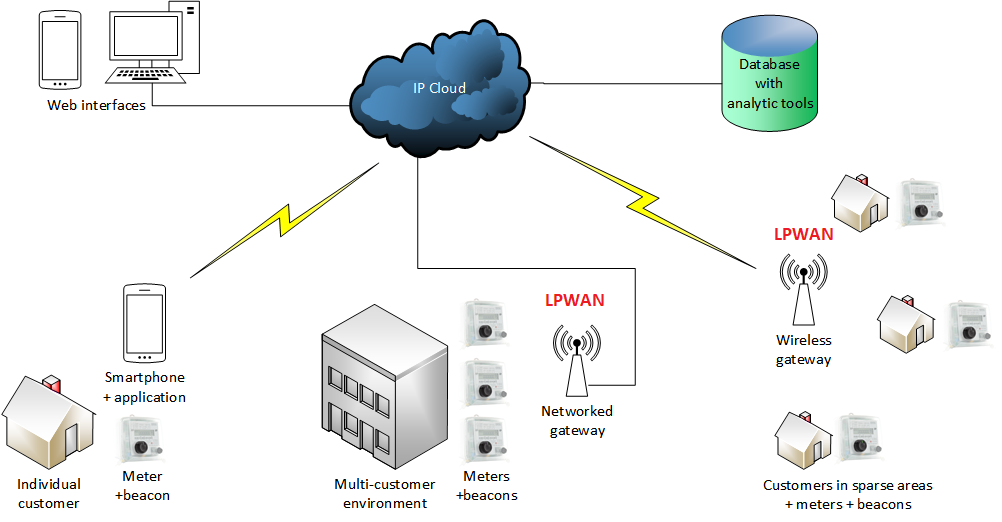}
\caption{Schematic diagram of the OneMeter 2.0 system. Potential areas of employing LPWAN protocols are marked in red.}
\label{fig:OMsystem}
\end{figure*}

According to the data of the European Commission~\cite{EU2023Monitoring}, so far, six EU members have achieved a full roll-out of smart meters: Denmark, Estonia, Finland, Italy, Spain, and Sweden. In 2022, France had about 92\% penetration, the Netherlands about 88\%, and Portugal 52\%, with full coverage expected by 2025. In Austria, Latvia, Poland, and the UK, the household penetration was significantly lower, with Austria at 47\%, Poland at 15\%, and the UK at 49\%. In the rest of the EU countries, the deployment of smart meters has varied significantly. 

This means, for example, in Poland, where there are 18 million metering points, only less than 2.7 million are equipped with smart meters. However, a remarkable part of the remaining electricity meters are equipped with optical ports, which are normally used for billing readouts but can be equally used to access the meter readouts using an optical sensor.

\subsection{Existing Solutions}
Several solutions exist that aim to acquire energy consumption data from existing electronic, non-smart meters. The Rhino Company offers the so-called RhinoAMI AP device~\cite{Rhino2018}, which accesses electronic meters via a DIN bus using a cable connection. Metering data can then be transmitted further using a GPRS or Ethernet connection. The device requires an external  5-12~V power source.

Smappee~\cite{Smappee2018} is another cable solution, offered currently at 229~EUR, which, in contrast to the previously described system, uses an electromagnetic sensor clipped to the phase cable supplying an electrical installation, e.g., in an apartment or an office. A dedicated application allows the monitoring of the current energy consumption. A proprietary Non-Intrusive Load Monitoring (NILM) algorithm helps to recognize individual electrical appliances. The Smappee metering system is powered by a 100-230~V main supply. It is noteworthy that Smappee, in fact, estimates the consumption instead of reading it from the meter.

A device called mReader{\textsuperscript\textregistered}Opto, produced by NUMERON~\cite{Numeron2018}, uses an optical sensor to communicate with the meter over the IEC~62056-21 protocol. It requires a USB connection to connect with a smartphone or a computer. It can work on a battery, but only for ca. 2h. The same producer also offers a gateway called smartBOX, which allows a remote transmission of the meter readouts over a network or GPRS.

REDZ Smart Communication Technologies offers another device with an optical sensor: KMK 118 Bluetooth Optical Probe~\cite{Redz2018}. Its functionality is similar to that of the previously described device, but here, cable communication is replaced by wireless communication. The device can be battery powered, but the battery life is reported to be only ``greater than 24h''. The device is offered at the price of 180 EUR.

\section{Our Solution}
\label{sec:OMProject}

\begin{table*}[!h]
	\caption{Comparison of LPWAN protocols~\cite{Raza2017LPWAN, Mekki2019Comparative}.}
	\small
	\begin{center}
		\begin{tabular}{|lcccccc|}
  %{|m{0.16\textwidth}|m{0.11\textwidth}|m{0.11\textwidth}|m{0.11\textwidth}|m{0.11\textwidth}|m{0.11\textwidth}|m{0.11\textwidth}|} 
  \hline
			Parameter & DASH7 & LoRa & LTE-M & NB-IoT & Sigfox & Weightless \\ \hline
			Bitrate [kb/s] & 9.6-166 & 0.3-37.5 & 1000 & 250 & 0.1 & 0.2-100 \\ 
			Data frame size [B] & 256 & 255 & 1000 & 1600 & 12 & 48 \\ 
			Daily limit [kB] & N/A & 3240 & \multicolumn{2}{c}{e.g., 50k (tariff-dependent)} & 1.68 & N/A \\ 
			Daily limit [messages] & N/A & 10 & \multicolumn{2}{c}{e.g., 500 (tariff-dependent)} & 140 & N/A \\ 
			Range urban/rural [km] & 1/5 & 2/15 & \multicolumn{2}{c}{1/10} & 10/40 & 2 \\ 
			Current in sleep-mode [mA] & bd. & 0.001 & 0.011 & 0.005 & 0.001 & N/A \\ 
			Max. current [mA] & N/A & 70 & 380 & 120 & 50 & N/A \\ 
			Regular activity & \multicolumn{2}{c}{required} & \multicolumn{2}{c}{non-required} &
			\multicolumn{2}{c|}{required} \\ 
			Corrective coding & \multicolumn{6}{c|}{yes} \\ \hline
		\end{tabular}
	\end{center}
	\label{tab:comparisionSpecs}
\end{table*}

We have developed a system that utilizes small, energy-efficient beacons with optical sensors to read data directly from electricity meters. In contrast to other existing solutions, our beacons are energy efficient, allowing them to work on a single battery for over a year. We propose employing either smartphones or dedicated gateways (e.g., LoRaWAN-based ones) to transfer measurement data to the cloud, as shown schematically in Figure~\ref{fig:OMsystem}. Thanks to a cloud-based data platform and the possibility of using user smartphones, our solution enables fast and cheap deployment of the AMI infrastructure using the existing, non-smart electricity meters. The details of the proposed solution are described below.

\subsection{Beacon with Optical Sensor}

A small bottle cap-shaped beacon of 32 mm diameter (compatible with the IEC~62056-21 interface) was designed, equipped with an optical sensor, LED diode, Nordic Semiconductor's processor nRF51, flash memory, Bluetooth Low Energy (BLE) radio components, and a 3.0V battery (CR2032 or double AA). The beacon is attached magnetically to an electronic meter equipped with an optical port. The optical sensor is designed with a miniature silicon photodiode of high radiant sensitivity and a low-power comparator. The optical sensor, together with the IR LED diode, are able to set up communication with a meter using the IEC~62056-21 (old: IEC~1107) or SML (Smart Message Language) protocol. The amount of measurement data acquired from the meter depends on the meter's model -- some of the meters present only the absolute active energy, while the others allow the readout of more detailed information, such as positive and negative active energy, or reactive energy. 

The processor was programmed in such a way that the beacon performs a readout of the meter every 15~min and stores the metering data in the flash memory. The BLE component allows other BLE devices to connect to the beacon to download metering data or to transmit the readout in real-time through BLE advertisement. 

\begin{figure}[!t]
\centering
\includegraphics[width=0.9\linewidth]{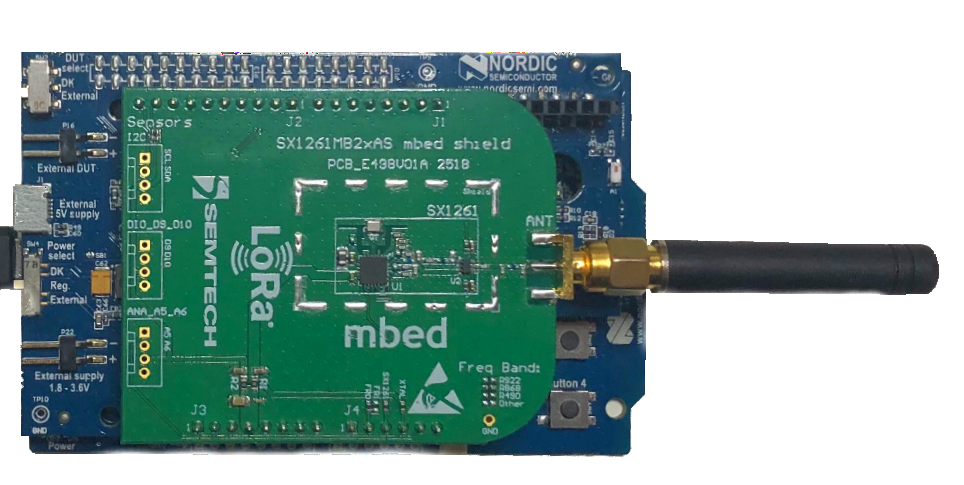}
\caption{Terminal unit, based on SX1261MB2BAS radio component, used for experiments with LoRaWAN transmission.}
\label{fig:LoRaUnit}
\end{figure}

\subsection{Data Platform}
The data platform provides gathering, analysis, and visualizations of the collected metering data. The platform was realized using the MongoDB database with a set of proprietary analytic algorithms. 

A web-based user interface allows the visualization of energy consumption data. The user is able to enter information about their tariff. The cost estimation of the consumed energy can be calculated thanks to the tariff data imported to the database for various energy re-sellers. The platform provides tools to generate reports showing consumption profiles for chosen date ranges and information about maximum power demand, including, for example, information on the percentage of time a certain power threshold was exceeded.

\subsection{Transmitting Measurement Data Using LPWAN Network}

While in our previous work~\cite{Brzozowski2019Enhancing}, we showed using smartphones as gateways to transmit measurement data from beacons to the cloud, in this study, we focus on using a dedicated gateway running an LPWAN protocol. Such an option may be used in urban areas to collect energy measurement data from multiple meters located, e.g., in a closed area or in a block of flats. It can also be advantageous in rural areas with less developed infrastructure, as depicted in~Figure~\ref{fig:OMsystem}.

Various LPWAN protocols were considered, such as DASH7, LoRa, LTE-M, SigFox, NB-IoT~\cite{Raza2017LPWAN}. Each of them has their advantages and drawbacks, as compared in~Table~\ref{tab:comparisionSpecs}. Considering the transmission range, the range is from $1$~km in urban areas
up to $40$~km in rural areas; the furthest are offered by LoRa and Sigfox. For the former, a spectacular record was achieved for the distance between
transmitter and receiver in favorable conditions: 702 km~\cite{telkamp17}.

Systems using LTE-derived standards do not require the installation of a radio gateway due to the existence of dedicated LTE infrastructure of mobile telephony. Sigfox standard is not officially supported in several countries (Poland included). Also, the LTE-M and NB-IoT infrastructure was insufficient in Poland at the time of running our experiments. Therefore, the cost of deploying communication links using these protocols would be very high. However, it is noteworthy that using LTE-M would be advantageous for prosumers and other advanced users, as this standard offers support for increased data transfer, which can be purchased from the telco provider. 

As for LoRa, its popularity is constantly growing. The advantage of the LoRa standard is the possibility of using free transmission, e.g., via The Things Network\footnote{\url{https://www.thethingsnetwork.org/}} or ChirpStack\footnote{\url{https://www.chirpstack.io/}}. Contrary to that, the cost of sending information using LTE-M or NB-IoT depends on the size of the packets. Considering the above factors and additional characteristics (e.g., security aspects), we chose the LoRaWAN standard for our experiments.

\section{Experimental setup}
\label{sec:Experiments}

\begin{figure}[t]
	\centering
	\includegraphics[height=15cm]{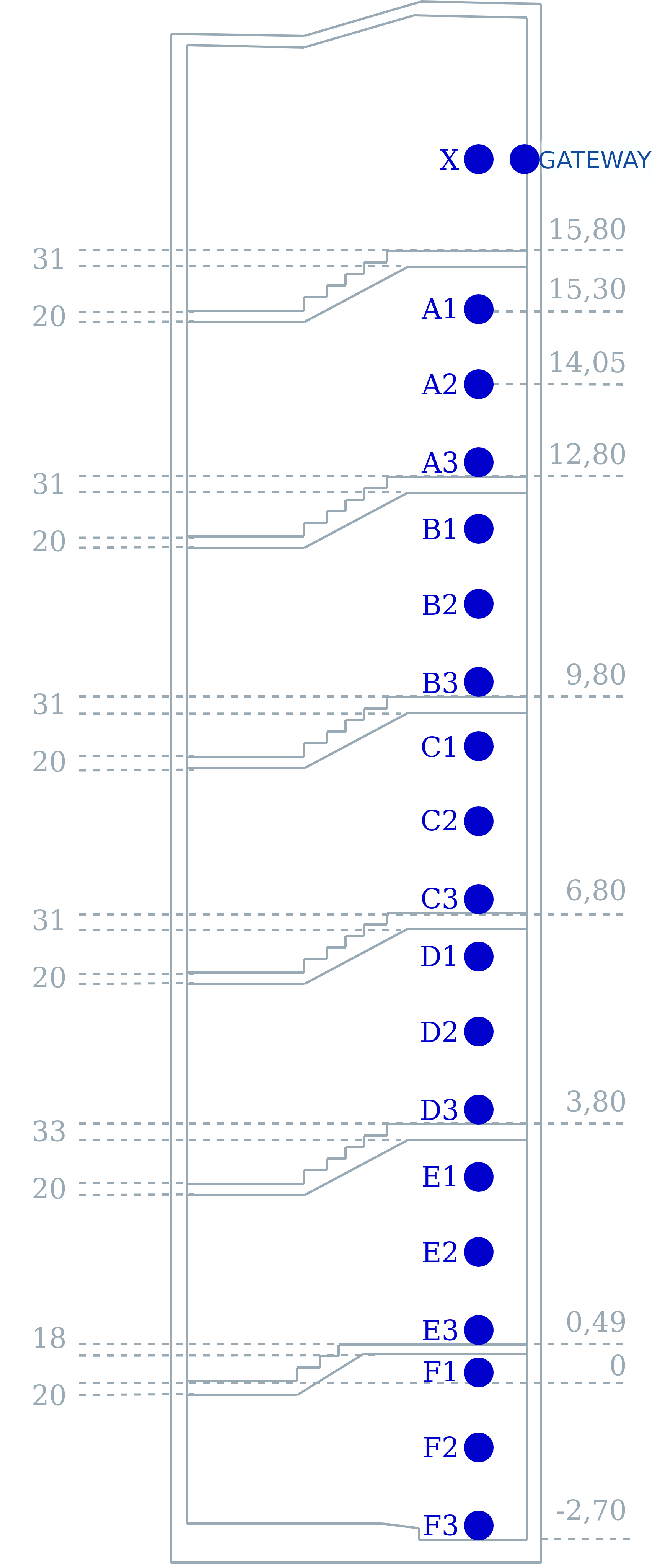}
	\caption{Location of the LoRaWAN gateway and the test points inside the building for experiments with indoor transmission. The values on the left indicate the thickness of the concrete elements [in centimeters]. The values on the right indicate the height relative to the ground [in meters].}
	\label{fig:LoRaIndoorMap}
\end{figure}

In the experiments described in this work, we planned to verify the energy consumption of the LoRaWAN-based transmission unit, the time required for transmission, and the coverage offered in real conditions. In this case, we focused on the urban environment.

As the base module, we used Nordic Semiconductor's nRF52 Development Kit 
 with nRF52832 processor. Its core element is the ARM Cortex M4  microcontroller with a $60$~MHz clock speed, $512$~kB flash memory, $64$~kB RAM memory, $32$~configurable I/O ports, and automatic processor power supply control system in the range of $1.7-3.6$~V. It is very energy-effective, its max. current should not exceed $8$~mA during CPU operations, $50$~µA in sleep mode, and $2$~µA in deep sleep mode.

 To enable radio transmission using LoRa protocol, we extended it with a radio component: SX1261MB2BAS device with the SX1261 processor (QFN24) from Semtech, with a radio frequency switch PE4259 from Peregrine Semiconductor and 14AC8253 antenna, as visualized in~Figure~\ref{fig:LoRaUnit}.
The nominal voltage of the module is $3.3$ V, and the transceiver is designed to operate in the non-commercial band in the voltage range $1.8–3.7$~V, taking into account sleep and standby modes to increase the module's energy savings. The maximum allowed transmission clock frequency is $16$~MHz. The crystal oscillator used (EXS00A-CS06465 $32$~MHz) meets the required frequency drift limitation at a level not higher than $\pm30$~ppm to ensure stable radio transmission. The SX1261 processor offers a maximum link budget of $163$~dB, transmitter power of $15$~dBm, and receiver sensitivity of $-137$~dBm. Considering the antenna's gain equal $2.15$~dBi, the equivalent isotropically radiated power (EIRP) will be $16.15$~dB, assuming transmission losses at the level of $1$~dB. 

As the LoRaWAN gateway, we chose The Things Gateway TTN-001-868-1.0, with LG8271--based radio transceiver. It offers data transmission with a power of up to $14$~dBm, as well as reception on eight transmission channels. The receiver sensitivity (for a bandwidth of $125$~kHz), according to the supplier documentation, was from $-126$~dBm to $-140$~dBm. For research purposes, we used The Things Network server, with The Things Stack toolkit.

When measuring the time and energy required by LoRaWAN transmission, we changed the payload size from $1$ to $50$B, which is the range of a typical payload with energy consumption data.

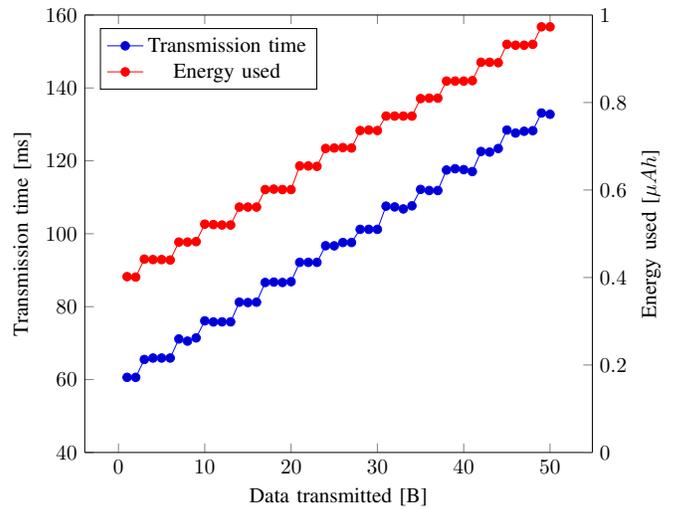
\begin{figure}[!t]
	\centering
	\begin{tikzpicture}[scale=0.8]
	\pgfplotsset{
		scale only axis
	}
	\begin{axis}[
	xlabel={Data transmitted [B]},
	ylabel={Transmission time [ms]},
	axis y line*=left,
	ymin=40, ymax=160
	]
	\addplot table [x=size, y=duration]{payload-cost-sf7.csv};\label{plot:LoRaDurationVsSizeSF7}
	\end{axis}
	\begin{axis}[
	ylabel={Energy used [$\mu Ah$]},
	axis y line*=right,
	axis x line=none,
	ymin=0, ymax=1,
	legend pos=north west
	]
	\addlegendimage{/pgfplots/refstyle=plot:LoRaDurationVsSizeSF7}\addlegendentry{Transmission time}
	\addplot+[red, mark options={red}] table [x=size, y=charge]{payload-cost-sf7.csv};
	\addlegendentry{Energy used};
	\end{axis}
	\end{tikzpicture}
	\caption{Time and energy required to transmit various amounts of data via LoRaWAN communication system, for low spreading factor. $P=13~dBm,~SF=7$.}
	\label{fig:LoRaPayloadCostSF7}
\end{figure}

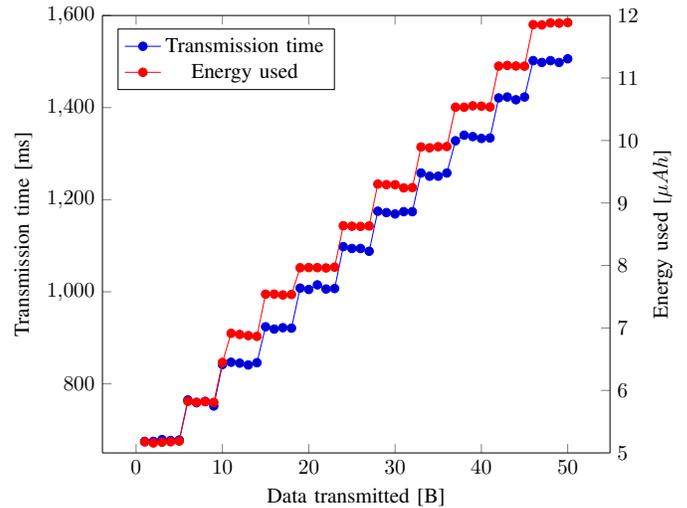
\begin{figure}[!t]
	\centering
	\begin{tikzpicture}[scale=0.8]
	\pgfplotsset{
		scale only axis
	}
	\begin{axis}[
	xlabel={Data transmitted [B]},
	ylabel={Transmission time [ms]},
	axis y line*=left,
	ymin=650, ymax=1600
	]
	\addplot table [x=size, y=duration]{payload-cost-sf11.csv};\label{plot:LoRaDurationVsSizeSF11}
	\end{axis}
	\begin{axis}[
	ylabel={Energy used [$\mu Ah$]},
	axis y line*=right,
	axis x line=none,
	ymin=5, ymax=12,
	legend pos=north west
	]
	\addlegendimage{/pgfplots/refstyle=plot:LoRaDurationVsSizeSF11}\addlegendentry{Transmission time}
	\addplot+[red, mark options={red}] table [x=size, y=charge]{payload-cost-sf11.csv};
	\addlegendentry{Energy used};
	\end{axis}
	\end{tikzpicture}
	\caption{Time and energy required to transmit various amounts of data via LoRaWAN communication system, for an increased spreading factor. $P=13~dBm,~SF=11$.}
	\label{fig:LoRaPayloadCostSF11}
\end{figure}

\begin{figure*}[!ht]
	\begin{center}
		\subfloat{
			\begin{tikzpicture}[scale=0.9]
			\begin{axis} [
			xtick=data,
			minor y tick num = 4,
			ymajorgrids,
			ymin=-120, ymax=-50,
			xlabel={Distance from gateway [m]},
			ylabel={RSSI [dBm]},
			xtick={0,3,6,9,12,15,18,21}
			]
			\addplot+[color=blue, mark=none, line width=2pt] coordinates {
				( 0.00,  -61.16)
				( 2.00,  -77.65)
				( 3.25,  -73.55)
				( 4.50,  -78.57)
				( 5.00,  -77.05)
				( 6.25,  -73.38)
				( 7.50,  -80.10)
				( 8.00,  -88.00)
				( 9.25,  -89.58)
				(10.50,  -87.72)
				(11.00,  -99.04)
				(12.25, -101.04)
				(13.50,  -99.58)
				(14.30, -108.94)
				(15.55, -105.80)
				(16.80, -103.98)
				(17.50, -110.80)
				(18.75, -109.90)
				(20.00, -112.10)
			}; \addlegendentry{$SF7$};
			\addplot+[color=red, mark=none, line width=2pt] coordinates {
				( 0.00,  -64.00)
				( 2.00,  -73.92)
				( 3.25,  -75.38)
				( 4.50,  -67.92)
				( 5.00,  -72.98)
				( 6.25,  -80.53)
				( 7.50,  -81.98)
				( 8.00,  -91.56)
				( 9.25,  -92.08)
				(10.50,  -91.00)
				(11.00,  -98.42)
				(12.25,  -99.55)
				(13.50, -101.00)
				(14.30, -105.82)
				(15.55, -108.14)
				(16.80, -109.56)
				(17.50, -110.65)
				(18.75, -111.70)
				(20.00, -113.09)
			}; \addlegendentry{$SF9$};
			\addplot+[color=gray, mark=none, line width=2pt] coordinates {
				( 0.00,  -62.58)
				( 2.00,  -70.78)
				( 3.25,  -78.52)
				( 4.50,  -74.36)
				( 5.00,  -78.12)
				( 6.25,  -81.24)
				( 7.50,  -81.94)
				( 8.00,  -86.76)
				( 9.25,  -88.38)
				(10.50,  -88.24)
				(11.00,  -97.00)
				(12.25,  -99.58)
				(13.50,  -95.86)
				(14.30, -106.24)
				(15.55, -105.20)
				(16.80, -104.84)
				(17.50, -110.04)
				(18.75, -109.72)
				(20.00, -111.46)
			}; \addlegendentry{$SF11$};
			\draw[dashed] (axis cs: 1.50, -50) -- (axis cs: 1.50, -120);
			\draw[dashed] (axis cs: 4.50, -50) -- (axis cs: 4.50, -120);
			\draw[dashed] (axis cs: 7.50, -50) -- (axis cs: 7.50, -120);
			\draw[dashed] (axis cs:10.50, -50) -- (axis cs:10.50, -120);
			\draw[dashed] (axis cs:13.50, -50) -- (axis cs:13.50, -120);
			\draw[dashed] (axis cs:16.80, -50) -- (axis cs:16.80, -120);
			\draw[dashed] (axis cs:20.00, -50) -- (axis cs:20.00, -120);
			\end{axis}
			\node (X) at (0.55, 6.00) {X};
			\node (A) at (1.45, 6.00) {A};
			\node (B) at (2.30, 6.00) {B};
			\node (C) at (3.15, 6.00) {C};
			\node (D) at (4.00, 6.00) {D};
			\node (E) at (4.85, 6.00) {E};
			\node (F) at (5.75, 6.00) {F};
			\end{tikzpicture}
		}
		\subfloat{
			\begin{tikzpicture}[scale=0.9]
			\begin{axis} [
			xtick=data,
			minor y tick num = 4,
			ymajorgrids,
			ymin=-4, ymax=12,
			xlabel={Distance from gateway [m]},
			ylabel={SNR [dB]},
			xtick={0,3,6,9,12,15,18,21}
			]
			\addplot+[color=blue, mark=none, line width=2pt] coordinates {
				( 0.00,  8.55)
				( 2.00,  8.91)
				( 3.25,  9.17)
				( 4.50,  9.29)
				( 5.00,  8.97)
				( 6.25,  9.19)
				( 7.50,  8.93)
				( 8.00,  8.92)
				( 9.25,  8.78)
				(10.50,  8.88)
				(11.00,  7.15)
				(12.25,  5.63)
				(13.50,  7.56)
				(14.30,  1.80)
				(15.55,  3.63)
				(16.80,  4.94)
				(17.50, -0.58)
				(18.75, -0.16)
				(20.00, -3.25)
			}; \addlegendentry{$SF7$};
			\addplot+[color=red, mark=none, line width=2pt] coordinates {
				( 0.00,  11.43)
				( 2.00,  11.29)
				( 3.25,  11.25)
				( 4.50,  11.14)
				( 5.00,  10.97)
				( 6.25,  11.03)
				( 7.50,  11.24)
				( 8.00,  10.42)
				( 9.25,  10.94)
				(10.50,  10.72)
				(11.00,   9.77)
				(12.25,   9.79)
				(13.50,   9.03)
				(14.30,  6.68)
				(15.55,  3.57)
				(16.80,  1.18)
				(17.50, -0.25)
				(18.75, -1.77)
				(20.00, -3.50)
			}; \addlegendentry{$SF9$};
			\addplot+[color=gray, mark=none, line width=2pt] coordinates {
				( 0.00,  10.85)
				( 2.00,  11.03)
				( 3.25,  10.49)
				( 4.50,  11.07)
				( 5.00,  10.90)
				( 6.25,  10.74)
				( 7.50,  10.56)
				( 8.00,  10.28)
				( 9.25,  10.69)
				(10.50,  10.48)
				(11.00,   8.88)
				(12.25,  8.45)
				(13.50,  9.81)
				(14.30,  5.55)
				(15.55,  5.66)
				(16.80,  6.04)
				(17.50, -0.29)
				(18.75,  0.61)
				(20.00, -2.08)
			}; \addlegendentry{$SF11$};
			\draw[dashed] (axis cs: 1.50, 15) -- (axis cs: 1.50, -4);
			\draw[dashed] (axis cs: 4.50, 15) -- (axis cs: 4.50, -4);
			\draw[dashed] (axis cs: 7.50, 15) -- (axis cs: 7.50, -4);
			\draw[dashed] (axis cs:10.50, 15) -- (axis cs:10.50, -4);
			\draw[dashed] (axis cs:13.50, 15) -- (axis cs:13.50, -4);
			\draw[dashed] (axis cs:16.80, 15) -- (axis cs:16.80, -4);
			\draw[dashed] (axis cs:20.00, 15) -- (axis cs:20.00, -4);
			\end{axis}
			\node (X) at (0.55, 6.00) {X};
			\node (A) at (1.45, 6.00) {A};
			\node (B) at (2.30, 6.00) {B};
			\node (C) at (3.15, 6.00) {C};
			\node (D) at (4.00, 6.00) {D};
			\node (E) at (4.85, 6.00) {E};
			\node (F) at (5.75, 6.00) {F};
			\end{tikzpicture}
		}
		\caption{Signal power (RSSI) and SNR against the distance from the LoRaWAN gateway inside a building during transmission between the beacon and the gateway, for various spreading factors (SF). Letters correspond to the consecutive floors (see~Figure~\ref{fig:LoRaIndoorMap}).}
		\label{fig:LoRaIndoorCoverage}
	\end{center}
\end{figure*}
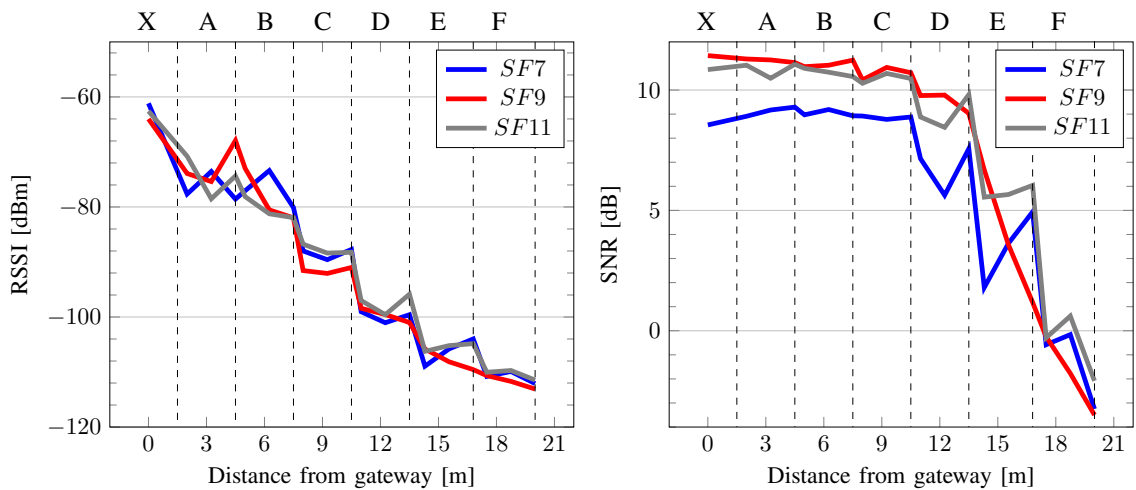

As for the experiments with LoRaWAN transmission range, we used two cases:
\begin{itemize}
    \item indoor transmission, when we measured the propagation of the radio signal within a multi-story building;
    \item outdoor transmission, when we measured the LoRaWAN coverage in the urban area.
\end{itemize}

For experiments with indoor transmission, we used a 6-floor building with a basement, made of reinforced concrete elements, with a LoRaWAN gateway installed on the top floor. A cross-section of the building is depicted in~Figure~\ref{fig:LoRaIndoorMap}. Such a setup is typical for collecting data from sensors installed on the electric meters, which are very often located in the staircase of a building. We measured the received LoRaWAN signal energy using the received signal strength indicator (RSSI) value, in dBm, and signal-to-noise ratio (SNR), measured in dB. We experimented with various spreading factor (SF) values to see if they had an impact on signal propagation and, as a consequence, on RSSI and SNR values.

When measuring the quality of the outdoor transmission, we kept $SF=7$, to consider the worst possible scenario. We measured the signal strength when moving the measuring terminal in the neighborhood of the building with the LoRaWAN gateway located in Warsaw, Poland. We used seven LoRaWAN transmission channels.

During the measurements, both indoor and outdoor, we used the transmission with bandwidth $BW=125~kHz$, code rate $CR=4/5$, preamble size of $8$~symbols, $2$B cyclic redundant code (CRC), and adaptive data rate (ADR) off.

\section{Results}
\label{sec:Results}

The minimum recorded current value (in sleep mode) was approximately $490$~µA. The obtained value is an order of magnitude higher than expected: we expected approx.~$50$~µA, considering the base and radio modules' energy demand in sleep mode. The probable cause was a software error, and the base module processor did not turn off some peripheral modules.
We also researched the relationship between the time and energy costs of message transmission for various payload size (see Figures~\ref{fig:LoRaPayloadCostSF7} and~\ref{fig:LoRaPayloadCostSF11}). The visible step functions suggest that the selection of the appropriate data size is important, i.e., when another threshold is exceeded, adding a few bytes does not result in an increase in cost transmission.

With a transfer of $50$~B for $SF = 7$ and $SF = 11$, approximately $19$~nAh and $240$~nAh of energy per byte were consumed, respectively.
Similarly, we observed $2.76$~ms and $30$~ms radio band usage time. Assuming the selected information payload size, transferring 3 kB of data per day would require $57$~µAh and $720$~µAh of energy and $8.28$~s and $90$~s of transmission time, respectively. Assuming a battery capacity of $1000$~mAh, its linear capacity decline, and no degradation cells (for rough estimation purposes only), the end device would be able to transmit data for almost $48$ and $4$ years, respectively.

The indoor range measurement results, shown in~Figure~\ref{fig:LoRaIndoorCoverage}, indicate no significant difference in signal quality when using different $SF$ values, which is most probably caused by the shape of the staircase. Of the values tested, the best results can be attributed to the $SF=9$ configuration. The $20$~m distance (i.e., 6~floors), including the $175$~cm-thick reinforced concrete ceiling, reduced the signal strength by~$50$~dBm and the SNR by around~$15$~dB. 

Considering these results, and also the gateway sensitivity (reported as being in the range from~$-140$~dBm to~$-126$~dBm), the intra-building LoRaWAN coverage in such types of buildings can be expected when the end device and the gateway are no more than 8-10 floors away. Therefore, for higher buildings of this type, it is recommended that the LoRaWAN gateways be placed in the middle of the building.

\begin{figure*}[!h]
	\centering
	\includegraphics[width=\linewidth]{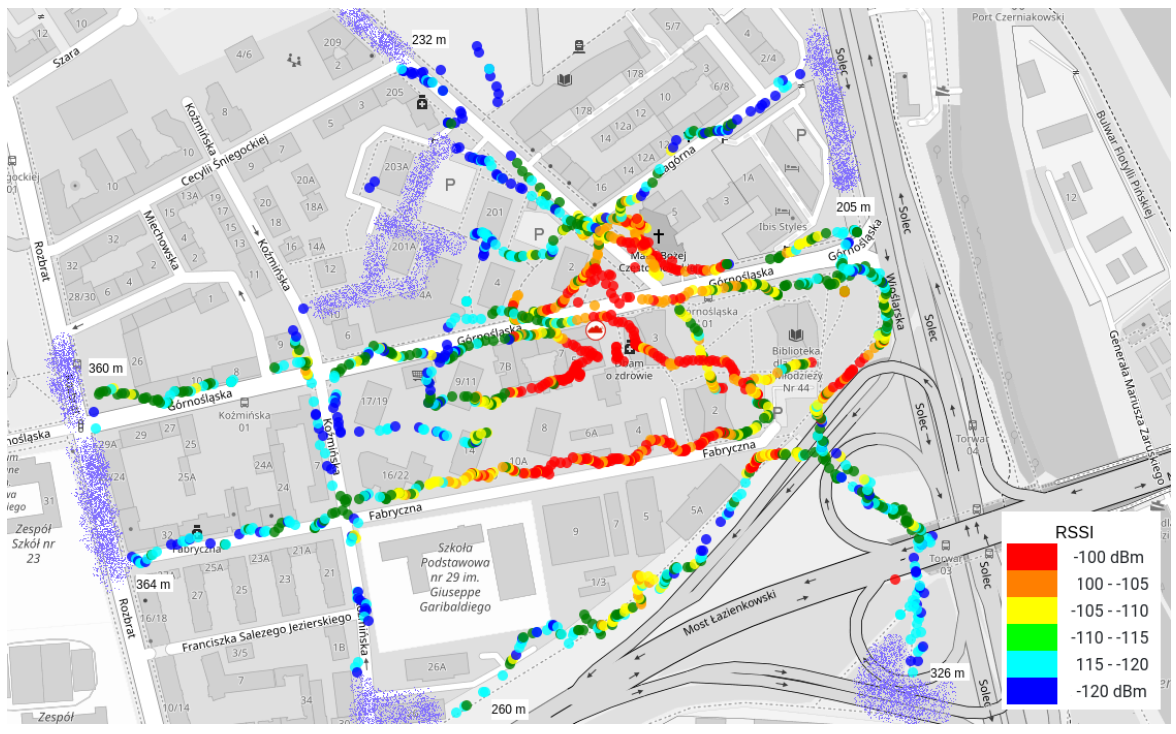}
	\caption{Outdoor LoRaWAN coverage in an urban dense area. A colored dot indicates the successful data transmission that has reached the gateway with the signal strength represented by the color. Purple blots mark places where several unsuccessful attempts were made to send data to the network. Distance labels denote the straight-line distances from the LoRa gateway. Coverage map by TheThingsNetwork.}
	\label{fig:LoRaOutdoorCoverage}
\end{figure*}

The results of outdoor measurements are depicted in~Figure~\ref{fig:LoRaOutdoorCoverage}. We observed that the maximum distance between nodes to ensure successful LoRaWAN data transmission was approx. $360$~m. We also observed that while the signal strength significantly decreased in the close vicinity of the gateway, it yielded RSSI
greater than $-100$~dBm along the long, straight streets. We think that this was the result of positive signal interference, which can be advantageous in future deployment. It must be remembered, however, that in our experiments, we used an indoor gateway, while in reality, an external unit will be used. Therefore, the coverage should be remarkably increased.

\section{Conclusion and Future Work}
\label{sec:Conclusions}

In this paper, we showed the results of experiments with using the LoRaWAN protocol for the transmission of energy measurement data. This research was a part of the OneMeter 2.0 project, which developed an electric energy measurement system based on beacons attached to regular, non-smart meters. 

In our study, we researched the energy consumption of a LoRa-based transmission unit, as well as the transmission range between network nodes in urban conditions. We discovered that LoRaWAN transmission was highly energy-efficient and offered decent coverage, even in a difficult, dense urban environment. Future work will involve experiments with transmission in rural areas, where the poor networking infrastructure will require radio LPWAN solutions, out of which LoRaWAN seems to be one of the best candidates.

% adjust the value as needed - may need to be readjusted if


%\bibliographystyle{IEEEtran}
% As suggested below, edit bibtemplate_samples.bib to reflect
% your bibliography. See bibtemplate.text for referencing.
%

\bibliographystyle{IEEEtran}
\bibliography{smartgrids}

\begin{thebibliography}{10}
\providecommand{\url}[1]{#1}
\csname url@samestyle\endcsname
\providecommand{\newblock}{\relax}
\providecommand{\bibinfo}[2]{#2}
\providecommand{\BIBentrySTDinterwordspacing}{\spaceskip=0pt\relax}
\providecommand{\BIBentryALTinterwordstretchfactor}{4}
\providecommand{\BIBentryALTinterwordspacing}{\spaceskip=\fontdimen2\font plus
\BIBentryALTinterwordstretchfactor\fontdimen3\font minus \fontdimen4\font\relax}
\providecommand{\BIBforeignlanguage}[2]{{%
\expandafter\ifx\csname l@#1\endcsname\relax
\typeout{** WARNING: IEEEtran.bst: No hyphenation pattern has been}%
\typeout{** loaded for the language `#1'. Using the pattern for}%
\typeout{** the default language instead.}%
\else
\language=\csname l@#1\endcsname
\fi
#2}}
\providecommand{\BIBdecl}{\relax}
\BIBdecl

\bibitem{EuDirective2012}
{European Union}, ``{Directive 2012/27/EU of the European Parliament and of the COUNCIL of 25 October 2012 on energy efficiency, amending Directives 2009/125/EC and 2010/30/EU and repealing Directives 2004/8/EC and 2006/32/EC},'' Official Journal of the European Union, Nov. 2012.

\bibitem{Morello2017Smart}
R.~Morello, C.~D. Capua, G.~Fulco, and S.~C. Mukhopadhyay, ``{A Smart Power Meter to Monitor Energy Flow in Smart Grids: The Role of Advanced Sensing and IoT in the Electric Grid of the Future},'' IEEE Sensors Journal, vol.~17, no.~23, Dec 2017, pp. 7828--7837.

\bibitem{Brzozowski2019Enhancing}
\BIBentryALTinterwordspacing
M.~Brzozowski, M.~Kruszewski, and A.~Janicki, ``Enhancing electricity meters with smart functionality using metering system with optical sensors,'' in Proc. the 4th International Conference on Advances in Sensors, Actuators, Metering and Sensing (ALLSENSORS 2019).\hskip 1em plus 0.5em minus 0.4em\relax IARIA Xpert Publishing Services, 2019, pp. 29--35. [Online]. Available: \url{https://api.semanticscholar.org/CorpusID:231743519}
\BIBentrySTDinterwordspacing

\bibitem{Garcia2017Power}
F.~D. Garcia, F.~P. Marafao, W.~A. de~Souza, and L.~C.~P. da~Silva, ``{Power Metering: History and Future Trends},'' in 9th IEEE Green Technologies Conference (GreenTech 2017), March 2017, pp. 26--33.

\bibitem{EU2023Monitoring}
{De Paola, A., Andreadou, N., Kotsakis, E., - Joint Research Centre (European Commission)}, ``{Clean Energy Technology Observatory: Smart Grids in the European Union - 2023 Status Report on Technology Development Trends, Value Chains and Markets},'' Oct 2023.

\bibitem{Rhino2018}
``{Rhino Energy - Solution - Hardware},'' https://rhino.energy/hardware/, 2018, accessed 08 Jan 2019.

\bibitem{Smappee2018}
``{Smappee Energy},'' https://www.smappee.com/be\_en/energy-monitor, 2018, accessed 08 Jan 2019.

\bibitem{Numeron2018}
NUMERON, ``Dokumentacja {NUMERON}, {mReader{\textsuperscript\textregistered}4},'' https://docs.numeron.pl/mreader4/, 2018, accessed 08 Jan 2019.

\bibitem{Redz2018}
{REDZ Smart Communication Technologies Ltd.}, ``{KMK 119 -- Bluetooth optical probe - Smart wireless auto protocols detection probe},'' http://probeformeters.com/brochure/REDZ\_KMK119\_Detailed\_Datasheet.pdf, 2018, accessed 08 Jan 2019.

\bibitem{Raza2017LPWAN}
U.~Raza, P.~Kulkarni, and M.~Sooriyabandara, ``Low power wide area networks: An overview,'' IEEE Communications Surveys \& Tutorials, vol.~19, no.~2, 2017, pp. 855--873.

\bibitem{Mekki2019Comparative}
\BIBentryALTinterwordspacing
K.~Mekki, E.~Bajic, F.~Chaxel, and F.~Meyer, ``A comparative study of lpwan technologies for large-scale iot deployment,'' ICT Express, vol.~5, no.~1, 2019, pp. 1--7. [Online]. Available: \url{https://www.sciencedirect.com/science/article/pii/S2405959517302953}
\BIBentrySTDinterwordspacing

\bibitem{telkamp17}
\BIBentryALTinterwordspacing
T.~Telkamp and L.~Slats, ``Ground breaking world record! lorawan packet received at 702~km distance.'' 2017, (Accessed on May 22, 2024). [Online]. Available: \url{https://www.thethingsnetwork.org/article/ground-breaking-world-record-lorawan-packet-received-at-702-km-\436-miles-distance}
\BIBentrySTDinterwordspacing

\end{thebibliography}

\end{document}